\documentclass[manuscript]{aastex}

\shorttitle{ALMA observations of a high density core in Taurus}
\shortauthors{Tokuda et al.}

\begin{document}


\title{ALMA OBSERVATIONS OF A HIGH-DENSITY CORE IN TAURUS: DYNAMICAL GAS INTERACTION AT THE POSSIBLE SITE OF A MULTIPLE STAR FORMATION}


\author{Kazuki Tokuda\altaffilmark{1}, Toshikazu Onishi\altaffilmark{1}, Kazuya Saigo\altaffilmark{2}, Akiko Kawamura\altaffilmark{2}, Yasuo Fukui\altaffilmark{3}, Tomoaki Matsumoto\altaffilmark{4}, Shu-ichiro Inutsuka\altaffilmark{3}, Masahiro N. Machida\altaffilmark{5}, Kengo Tomida\altaffilmark{6,7}, AND Kengo Tachihara\altaffilmark{3}}

%\affil{Space Telescope Science Institute, Baltimore, MD 21218}

\email{}


\altaffiltext{1}{Department of Physical Science, Graduate School of Science, Osaka Prefecture University, 1-1 Gakuen-cho, Naka-ku, Sakai, Osaka 599-8531, Japan; s\_k.tokuda@p.s.osakafu-u.ac.jp}
\altaffiltext{2}{National Astronomical Observatory of Japan, Mitaka, Tokyo 181-8588, Japan}
\altaffiltext{3}{Department of Physics, Nagoya University, Chikusa-ku, Nagoya 464-8602, Japan}
\altaffiltext{4}{Faculty of Humanity and Environment, Hosei University, Fujimi, Chiyoda-ku, Tokyo 102-8160, Japan}
\altaffiltext{5}{Department of Earth and Planetary Sciences, Kyushu University, Fukuoka 812-8581, Japan}
\altaffiltext{6}{Department of Astrophysical Sciences, Princeton University, Princeton, NJ 08544, USA}
\altaffiltext{7}{Department of Physics, The University of Tokyo, Tokyo 113-0033, Japan}


\begin{abstract}
Starless dense cores eventually collapse dynamically, which forms protostars inside them, and the physical properties of the cores determine the nature of the forming protostars.  We report ALMA observations of dust continuum emission and molecular rotational lines toward MC27 or L1521F, which is considered to be very close to the first protostellar core phase. 
We found a few starless high-density cores, one of which has a very high density of $\sim$10$^{7}$cm$^{-3}$, within a region of several hundred AU around a very low-luminosity protostar detected by {\em Spitzer}.  
A very compact bipolar outflow with a dynamical timescale of a few hundred years was found toward the protostar.  
The molecular line observation shows several cores with an arc-like structure, possibly due to the dynamical gas interaction.  These complex structures revealed in the present observations suggest that the initial condition of star formation is highly dynamical in nature, which is considered to be a key factor in understanding fundamental issues of star formation such as the formation of multiple stars and the origin of the initial mass function of stars.
\end{abstract}



\section{Introduction}

In low-mass star forming regions, most of the starless dense cores, whose densities are 10$^{4-6}$ cm$^{-3}$, do not seem to be dynamically collapsing because the timescale statistically derived is larger than the free fall time \citep{Onishi2002} and because the density distribution is flat toward the peak \citep{Ward1994}.  The subsequent evolutionary stage we can easily observe is a Class 0/I protostar phase, which shows strong outflow, indicating that the gravity is strong, and then mature protostellar cores were already formed. 

Recent observations of Class 0 protostars, the earliest phase of protostars whose ages are 10$^{4-5}$ yr, in nearby low-mass star forming regions have revealed a high frequency ($\sim$2/3) of binary or multiple protostars \citep{Maury2010,Chen2013}, and complex gas structures in the envelopes on spatial scales from 0.1 pc to $\sim$1000 AU \citep{Tobin2010,Tobin2011}.  However, there remains a missing observational link between the complex envelope and formed protostars.  It is key in the understanding of the origin of binary or multiple systems, for example, how the surrounding gas interacts with each other to form stars at the center of a dense core. It has been difficult to observe the objects between the starless and Class 0/I stages because the timescale is very short and the identification is observationally difficult.  

Recently, there have been many studies for the objects containing the first protostellar core \citep{Onishi1999,Enoch2010,Chen2012}.  One of such objects is MC27 \citep{Mizuno1994,Onishi1996,Onishi1998,Onishi1999,Onishi2002} or L1521F \citep{Codella1997}.
It is of very high density and contains one of the faintest protostars detected by {\em Spitzer} (hereafter,  the {\em Spitzer} source), which indicates that it is among the youngest and still may preserve the initial conditions of star formation \citep{Bourke2006,Terebey2009}.
\citet{Bourke2006} suggests that the {\em Spitzer} source lies at the apex of a bright but compact conical-shaped nebula at 4.5 $\mu$m, which opens to the west, and \citet{Terebey2009} also suggests that the source has a moderate inclination.
Previous observations found that the spatial distribution of MC27 in three different density tracers ranging from 10$^{3-5}$ cm$^{-3}$ is consistent with steep power-law distribution of an index of $-1.9$ \citep{Onishi1999}, which is consistent with the infalling envelope of the Larson--Penston solution in the early phase of star formation, and is different from what is usually derived for several starless cores \citep{Ward1994}.  An indication of compact and dense molecular outflow whose size is smaller than 1500 AU is also detected \citep{Onishi1999}.  Interferometric dust observation detected a single source \citep{Maury2010,Chen2013}, which coincides with the protostar, and no other features were detected although the surrounding envelope has complex velocity structures as revealed by molecular line observations \citep{Tobin2011}.  These interferometric observations with high angular resolutions failed to detect extended high-density gas around the {\em Spitzer} source due to the lack of the sensitivity and to the narrow spatial frequency coverage.

\section{Observations}

We carried out ALMA Band 6 (211--275 GHz) observations toward MC27/L1521F with the center position of ($\alpha_{J2000.0}$, $\delta_{J2000.0}$) = (4$^{\rm h}$28$^{\rm m}$38\fs92, +26\arcdeg51\arcsec36\farcs2) during Early Science Cycle 0.  The target molecular lines were HCO$^+$ ($J$ = 3--2; 267.55763 GHz), HCN ($J$ = 3--2; 265.88643 GHz), H$^{13}$CO$^+$ ($J$ = 3--2; 260.25534 GHz), CS ($J$ = 5--4; 244.93556 GHz), and SiO ($J$ = 6--5; 260.51802 GHz).  For the first two lines, seven-pointing mosaic observations have been carried out around the center position on 2012 January 11 and on 2012 November 5 and 6, and single-pointing observations for the rest on 2012 January 11 and on 2012 November 5.  The January observations were carried out by using 16 antennas with a compact configuration, and the November observations by 24 antennas with a compact+extended configuration.  The baselines range from 26.5 m to 381.8 m, corresponding to {\em uv} distances from 12.2 to 211 k$\lambda$.  The correlator was used in the frequency domain mode with a bandwidth of 234.4 MHz (61.04 kHz $\times$ 3840 channels).  
The calibration of the complex gains was carried out through observations of J0423--013, phase calibration on 3c111, and flux calibration on J0538--440, Callisto and Ganymede.  For the flux calibration of the solar system objects, we used the Butler-JPL-Horizons 2012 model (https://science.nrao.edu/facilities/alma/aboutALMA/Technology/ALMA\_Memo\_Series/\\alma594/abs594).
The data were processed in the Common Astronomy Software Applications package (http://casa.nrao.edu), and visibilities imaged.  The synthesized beam is $\sim$1\farcs0 $\times$ 0\farcs8 = 140 AU $\times$ 110 AU for HCO$^+$ ($J$ = 3--2) and scales approximately with the frequency for other transitions (see also Figures \ref{fig1}--\ref{fig3}).  The (1$\sigma$) rms of molecular lines are $\sim$8 mJy beam$^{-1}$ with a velocity resolution of $\sim$0.14 km s$^{-1}$.  The dust continuum emission was extracted by integrating all the frequency ranges except for the line frequencies. The synthesized beam and rms for the continuum map are $\sim$1\farcs1 $\times$ 0\farcs8, and 0.12 mJy beam$^{-1}$, respectively.

\section{Results and Discussions}

\subsection{High-density Dust Cores} \label{HDDC}

\begin{figure}
\epsscale{.99}
\plotone{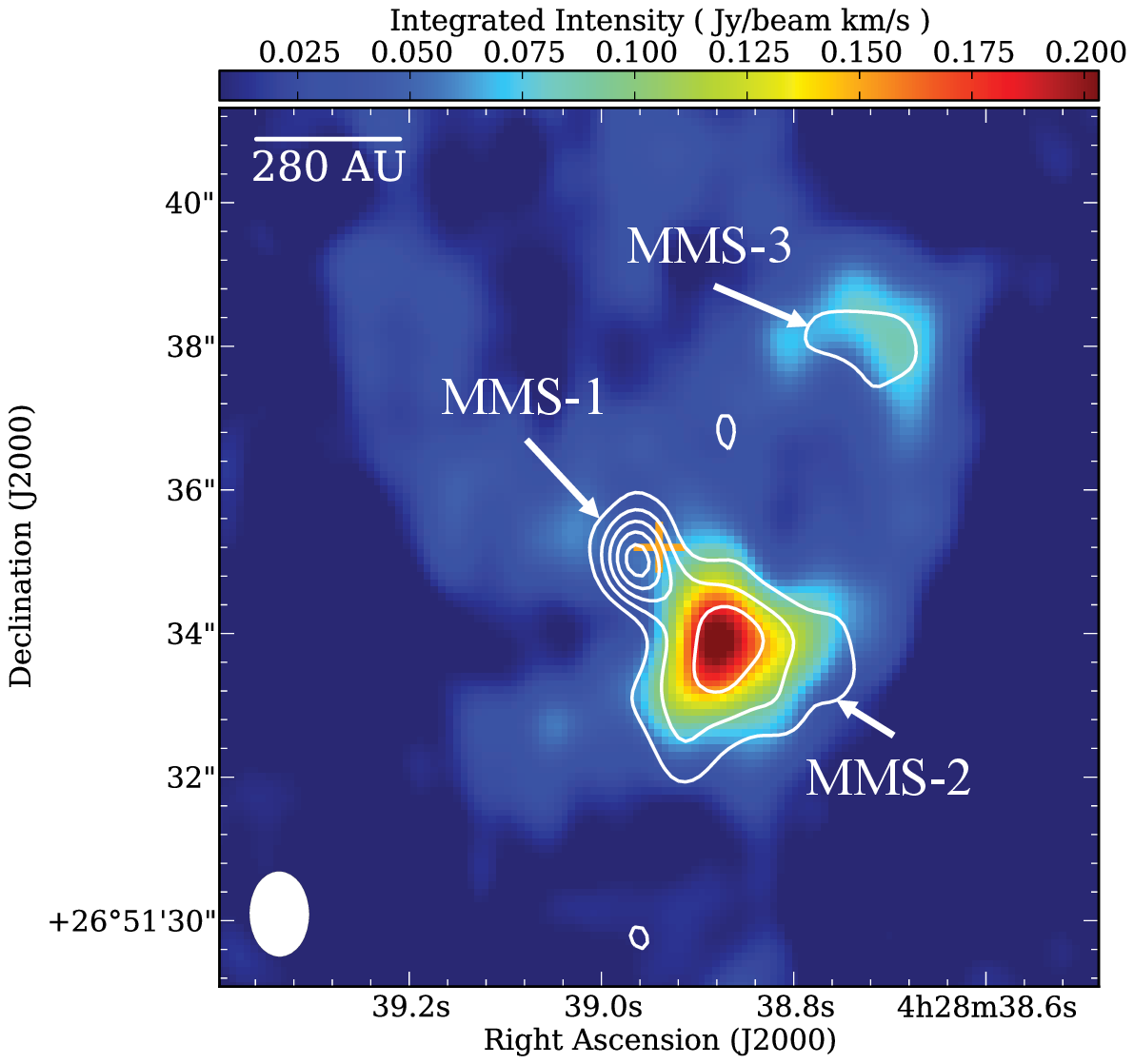}
\caption{Images of the high-density condensations of MC27/L1521F. 
ALMA images of 1.1 mm dust continuum emission in white contours and total velocity integrated intensity of H$^{13}$CO$^+$ ($J$ = 3--2) in color scale with the velocity  range of $5.0$--$8.0$ km s$^{-1}$. The contours start at three times the noise level and increase at this interval; the noise level is 0.12 mJy beam$^{-1}$.  Orange cross indicates the position of the {\em Spitzer} source, which also coincides with that of 1.3 mm continuum peak of the PdBI observation \citep{Maury2010}.  The angular resolution of the H$^{13}$CO$^+$ ($J$ = 3--2) is given by the ellipse in the lower left corner, 1\farcs2 $\times$ 0\farcs8. \label{fig1}}
\end{figure}

 Figure \ref{fig1} shows that there are three major continuum peaks (the parameters are listed in Table \ref{table1}), and one of which, MMS-1, coincides with the position of the {\em Spitzer} source.  MMS-1 is not spatially resolved by the present observation with an angular resolution of $\sim$1\arcsec, and was also detected as a point source with the Plateau de Bure Interferometer (PdBI) observation with an angular resolution of $\sim$0\farcs4 \citep{Maury2010}.  We have not detected any line intensity enhancements toward the {\em Spitzer} source, confirming that the {\em Spitzer} source has no associated extended envelope.  

Figure \ref{fig1} also shows that there is an extended core, MMS-2, with a beam-convolved diameter of $\sim$380 AU (see Table \ref{table1}) to the southwest of the {\em Spitzer} source.  A clear intensity enhancement of H$^{13}$CO$^+$($J$ = 3--2), a high-density gas tracer, is seen toward MMS-2, showing that the dust continuum peak is due to the density enhancement.

We calculated the mass from the dust emission assuming optically thin emission and constant dust temperature with the following equations; $M = F_\nu d^2/\kappa_\nu B_ \nu(T_d)$ and $\kappa_\nu = \kappa_{231{\rm GHz}}(\nu /231 {\rm GHz})^{\beta}$ where, $\kappa_\nu$ is the mass absorption coefficient, $B_{\nu}$ is the Plank function, $F_\nu$ is the integrated flux of the continuum emission at frequency $\nu$, $T_d$ is the dust temperature, $d$ is the distance of the source (140 pc \citep{Elias1978}), $\kappa_{\rm 231GHz}$ = 0.01 cm$^2$ g$^{-1}$ of interstellar matter is the emissivity of the dust continuum at 231 GHz \citep{Oss1994,Kau2008}, and $\beta$ is the dust emissivity index.  We also estimated the column density of H$_{2}$ from the dust emission with the following equation; $N = F_{\nu}^{\rm beam}/\Omega_{\rm A} \mu_{\rm H_{2}}m_{\rm H} \kappa_{\nu} B_{\nu}(T)$, where $F_{\nu}^{\rm beam}$ is the flux per beam, $\Omega_{\rm A}$ is the solid angle of the beam, $\mu_{\rm H_{2}}$ is the molecular weight per hydrogen, and $m_{\rm H}$ is the H-atom mass. We assumed $T_d$ = 10 K and $\beta$ = 1 \citep{Kau2008}.  The derived parameters are shown in Table \ref{table1}. 

Under the assumption of the local thermodynamical equilibrium condition and optically thin molecular line emission, we also estimated the mass and column density from the H$^{13}$CO$^+$ ($J$ = 3--2) data. The excitation temperature is adopted as 10 K. The fractional abundance of H$^{13}$CO$^+$ relative to H$_2$ is adopted as $X_{\rm H^{13}CO^{+}}$ = 1.0 $\times$ 10$^{-9}$ \citep{But1995}. The derived parameters are shown in Table \ref{table1}. 

The density of MMS-2 derived by dividing the dust column density by the size is calculated to be $\sim$10$^7$ cm$^{-3}$, which is so far the highest among starless gas condensations in low-mass star forming regions. The peak antenna temperature of the H$^{13}$CO$^+$ ($J$ = 3--2) spectra is as high as $\sim$7 K, which means that the line gets almost optically thick and that we need observations in more rare species to look into the central part.  

These results indicate that this core is one of the most evolved starless condensations, or may contain very faint protostar of $<$ 0.1 $L_\sun$ not detected by {\em Spitzer}, which is a strong candidate for searching for a first protostellar core, the first quasi-hydrostatic object during the star formation process \citep{Larson1969,Masunaga2000,Tomida2013}.  

There is also another low-intensity dust core toward the northwest of the {\em Spitzer} source, MMS-3. There is an enhancement of H$^{13}$CO$^+$ ($J$ = 3--2) toward MMS-3, and this fact indicates that MMS-3 is also a high-density core with a density of 10$^{6-7}$ cm$^{-3}$.

\subsection{Very Compact Outflow} \label{VCO}

\begin{figure}
\epsscale{.70}
\plotone{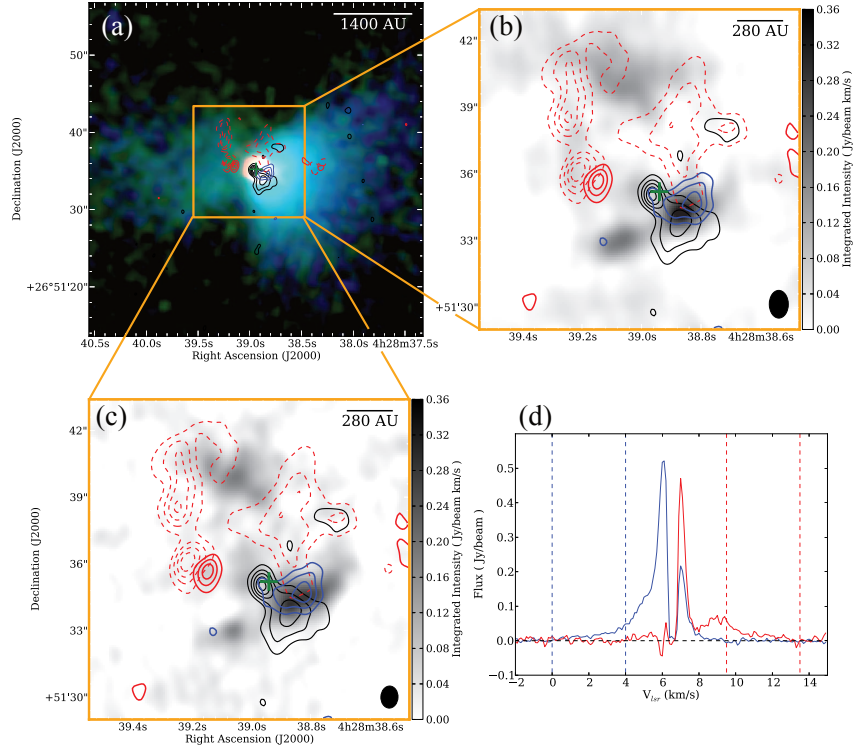}
\caption{Distribution of bipolar outflow emitted from the {\em Spitzer} source.
(a) Color is the R(8 $\mu$m)--G(4.5 $\mu$m)--B(3.5 $\mu$m) image with {\em Spitzer}. 
(b) Gray scale image shows total velocity-integrated intensities of CS ($J$ = 5--4). Red and blue solid contours show images of the velocity-integrated intensity of HCO$^+$($J$ = 3--2) with ranges of $9.0$--$13.5$ km s$^{-1}$ and $0.0$--$4.0$ km s$^{-1}$, respectively.  Red dashed contours show an image of velocity-integrated intensity of HCO$^{+}$ ($J$ = 3--2) with a range of $7.5$--$8.0$ km s$^{-1}$.  The lowest contour and subsequent contour step are 0.032 Jy beam$^{-1}$ km s$^{-1}$.  Black contours show the image of 1.1 mm dust continuum emission same as in Figure \ref{fig1}. The angular resolution of the CS ($J$ = 5--4) is given by the ellipse in the lower right corner, 1\farcs3 $\times$ 0\farcs9. The green cross is the position of the {\em Spitzer} source. 
(c) Gray scale image shows total velocity-integrated intensities of HCN ($J$ = 3--2).  Contours and the cross are same as panel (b). The angular resolution of the HCN ($J$ = 3--2) is given by the ellipse in the lower right corner, 1\farcs0 $\times$ 0\farcs8.
(d) Red and blue profiles show averaged spectra of the bipolar outflow over the regions inside the red and blue solid lowest contours, respectively, in panels (a)--(c).  The red and blue broken lines show the velocity ranges for the red and blue solid contours, respectively.
\label{fig2}}
\end{figure}

\begin{figure}
\epsscale{.99}
\plotone{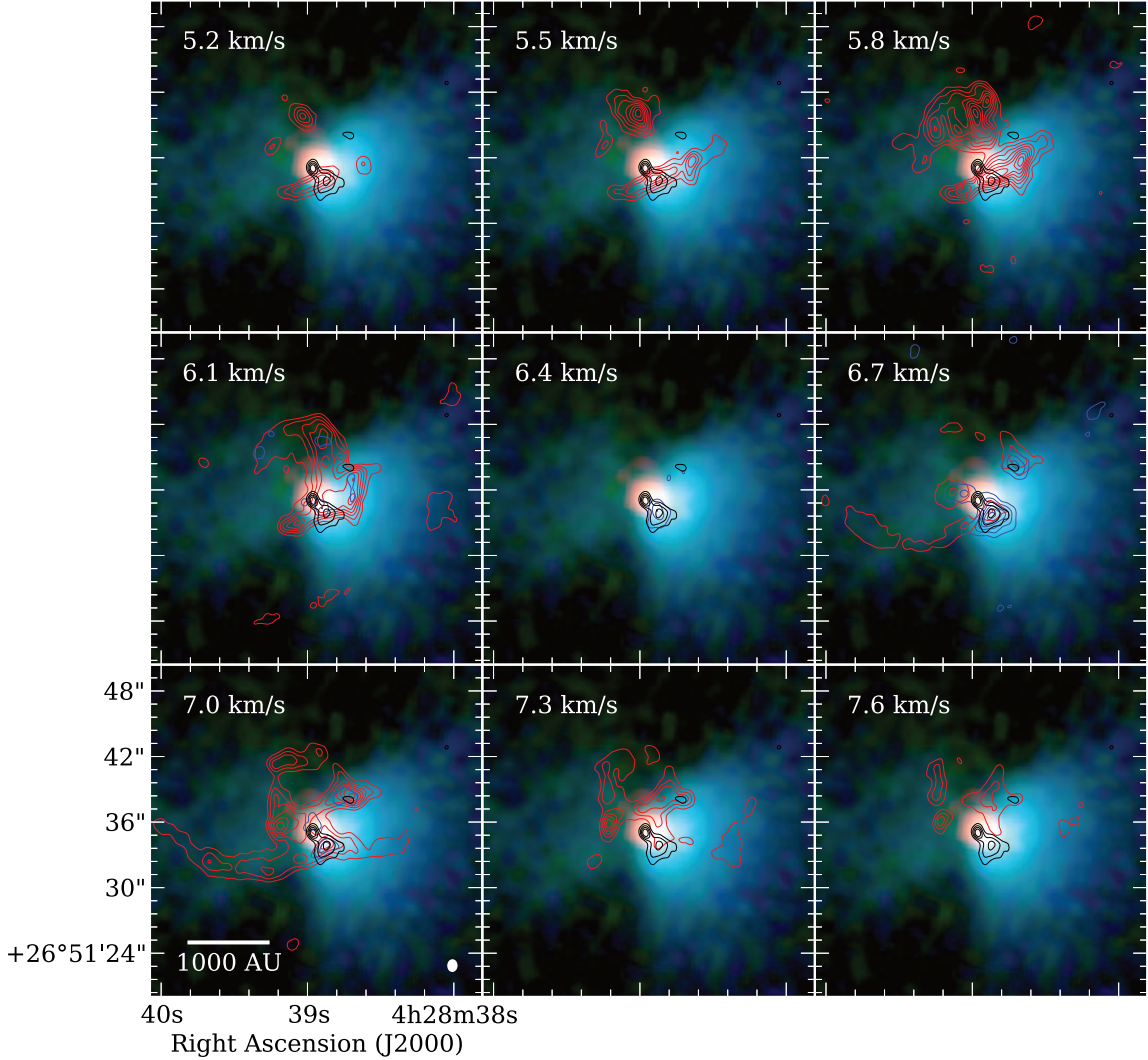}
\caption{Velocity channel maps of HCO$^+$($J$ = 3--2) and H$^{13}$CO$^{+}$ ($J$ = 3--2) emission toward MC27/L1521F.
The red--green--blue image shows the {\em Spitzer} observations same as in Figure \ref{fig2}(a).
Red and Blue contours show velocity-range-integrated intensity maps of HCO$^+$($J$ = 3--2) and H$^{13}$CO$^{+}$ ($J$ = 3--2) emission, respectively.  For HCO$^{+}$ ($J$ = 3--2), the lowest contour and subsequent step contour are 0.024 Jy beam$^{-1}$ km s$^{-1}$. For H$^{13}$CO$^{+}$ ($J$ =  3--2), they are 0.015 Jy beam$^{-1}$ km s$^{-1}$. The velocity width for each map is 0.3 km s$^{-1}$. The lowest velocities are given in upper left corner of each panel. The black contours show 1.1 mm dust continuum emission. The lowest contour and subsequent contour step are 0.48 and 0.36 mJy beam$^{-1}$, respectively.  The angular resolution of the HCO$^+$($J$ = 3--2) is given by the white ellipse in lower left corner of the bottom left panel, 1\farcs0 $\times$ 0\farcs8.
\label{fig3}}
\end{figure}

 We found a very compact bipolar outflow centered at the {\em Spitzer} source.  Figures \ref{fig2}(a)--(c) show the distribution of outflowing gas in HCO$^+$ ($J$ = 3--2) on the {\em Spitzer} image. Figure \ref{fig2}(d) shows the spectra toward the outflowing gas. Blueshifted gas is seen toward the west of the {\em Spitzer} source, and the redshifted gas toward the opposite direction.  
This direction is consistent with that of the outflow cavity; the west cavity toward us and the east cavity away from us \citep{Bourke2006}. \citet{Takahashi2013} claimed the first detection of a compact outflow in $^{12}$CO($J$ = 2--1) toward the {\em Spitzer} source, although the $^{12}$CO($J$ = 2--1) outflow direction is opposite to what we observed.  This may be because the critical density for the excitation of the $^{12}$CO($J$ = 2--1) is much lower than that of HCO$^{+}$($J$ = 3--2), and thus the $^{12}$CO($J$ = 2--1) just traces the velocity structure of the lower-density surrounding gas.

We estimated the size ($R_{\rm flow}$) and velocity ($V_{\rm flow}$) of the outflow for the red and blue components by use of the following equations; $R_{\rm obs} = R_{\rm flow} {\rm cos}i$, $V_{\rm obs} = V_{\rm flow} {\rm sin}i$,  where $R_{\rm obs}$ is the projected distance to the peak intensity of the outflow lobe position from the {\em Spitzer} source, and $V_{\rm obs}$ is the maximum radial velocity measured from HCO$^{+}$($J$ = 3--2) data.  Then, the dynamical timescale is calculated to be $R_{\rm flow}/V_{\rm flow}$. We assume here that the inclination angle $i$, which is defined as 0$\arcdeg$ for the plane of the sky, is $20$$\arcdeg$--$40$$\arcdeg$ \citep{Terebey2009}. Maximum velocities of blueshifted and redshifted components are $10$--$19$ and $11$--$20$ km s$^{-1}$, and the sizes are $260$--$310$ and $400$--$490$ AU, respectively. The dynamical timescale of the outflow is thus estimated to be a few hundred years (blue = $60$--$140$ yr, red =$90$--$210$ yr), and this indicates that mass accretion is still taking place at the {\em Spitzer} source.
 
Figures \ref{fig2}(a)--(c) also show that both the red- and blueshifted outflowing gas seem to delineate the gas with lower velocities; the blueshifted gas delineates the starless dust core (MMS-2), and the redshifted gas is located at the edge of the $7.5$--$8.0$ km s$^{-1}$ velocity component of HCO$^+$ ($J$ = 3--2).  This fact implies that the outflowing gas is interacting with the surrounding gas, and that the compression and/or abundance enhancement due to the shock are shown as the molecular outflow.  Figures \ref{fig2}(b) and (c) show the intensity distributions of the CS ($J$ = 5--4) and HCN ($J$ = 3--2), which are optically thicker than H$^{13}$CO$^+$ ($J$ = 3--2) and optically thinner than HCO$^{+}$ ($J$ = 3--2); the intensities have the peak close to the blueshifted outflow component rather than toward the H$^{13}$CO$^{+}$ ($J$ = 3--2) peak, implying that the molecular abundances are enhanced toward the interacting region. 
The reason why the redshifted outflow component is not continuously distributed from the source position can be explained that there was no dense gas between the source and the present redshifted component or the mass outflow is episodic.

We note that no SiO ($J$ = 6--5) emission was detected, which means that there is no strong shock to excite the line and is consistent with the fact that the {\em Spitzer} source is very young.

\subsection{Arc-like Structure} \label{Arc}

\begin{figure}
\epsscale{.99}
\plotone{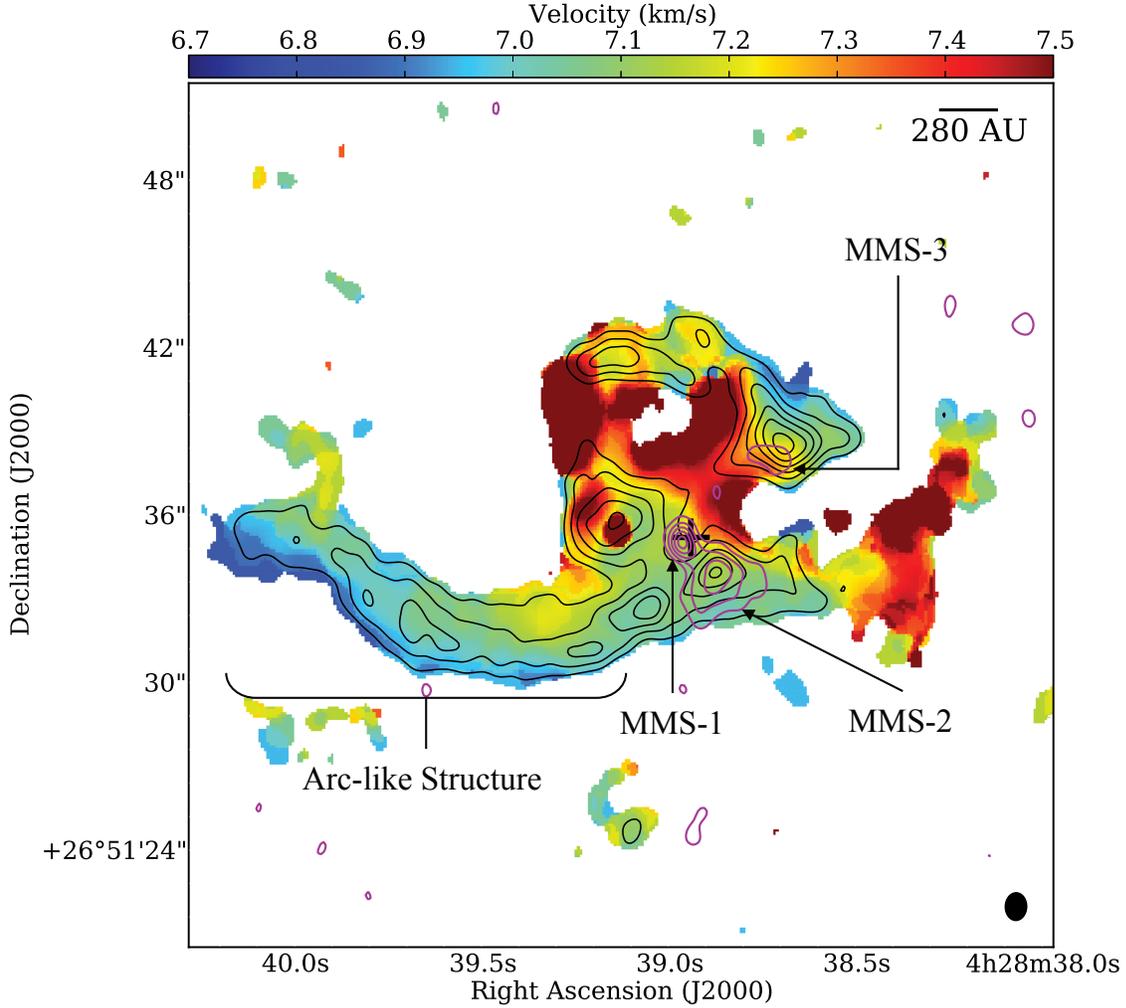}
\caption{Images of HCO$^{+}$ ($J$ = 3--2) emission of a red velocity component toward MC27/L1521F.
Black contour shows an image of velocity-integrated intensity of HCO$^+$($J$ = 3--2) with a velocity range of $6.8$--$7.0$ km s$^{-1}$. Both the lowest contour and subsequent contour step are 0.020 Jy beam$^{-1}$ km s$^{-1}$.  The first-moment intensity-weighted velocity map of the red velocity component with a range of $6.5$--$14.0$ km s$^{-1}$ is shown in color scale. The magenta contour is the image of dust continuum emission with the same contour levels as Figure \ref{fig1}. The black cross is the position of the {\em Spitzer} source. The angular resolution of the HCO$^+$($J$ = 3--2) is given by the ellipse in the lower right corner, 1\farcs0 $\times$ 0\farcs8.\label{fig4}}
\end{figure}

Figure \ref{fig3} shows the velocity channel maps of the HCO$^+$ ($J$ = 3--2) and H$^{13}$CO$^{+}$ ($J$ = 3--2) emission.
The systemic velocity of MC27/L1521F with single-dish observations is about 6.5 km s$^{-1}$ \citep[e.g.,][]{Onishi1999}, and we can observe no significant emission of HCO$^{+}$ ($J$ = 3--2) around the velocity. The lack of the emission is due to a combination of the optical thickness of the line and the extended emission missing due to the interferometry observation.  The striking feature is seen in the red component around 7 km s$^{-1}$; we see a long arc-like structure with a few core-like features (Figure \ref{fig4}).  The length of the arc-like structure is $\sim$2000 AU, and there is a slight velocity shift along and across the arc. The arc-like structure of HCO$^+$($J$ = 3--2) on 1000 AU scale can be considered as a consequence of the dynamical interaction between the small dense cores and the surrounding gas on that scale.  
The typical velocity of the arc-like structure is $\sim$0.5 km s$^{-1}$ with respect to the systemic velocity, and it is comparable to the dynamical velocity on that scale, e.g., the Kepler velocity for a mass of 0.1 $M_\sun$ and a radius of 1000 AU is 0.3 km s$^{-1}$.

Similar arc-like structures have been reported in the previous numerical simulations of the turbulent fragmentation models \citep{Bate2002,Goodwin2004,Offner2008}, where turbulence promotes fragmentation of cloud cores during the collapse and dynamical interaction of the fragments provide arcs or spiral arms in the envelopes.  This indicates that in MC27 the turbulence plays an important role in undergoing fragmentation in the central part of the cloud core, which is different from the classic scenarios of fragmentation in massive disks \citep{Larson1987,Boss2002,Machida2008}.  The driving source of the turbulence may be the complex velocity structure in the larger scale surrounding gas observed in the previous observation \citep{Tobin2011}.  A possible scenario is an interaction of cloud cores, which are formed by the fragmentation of a filamentary cloud.  The filamentary structures in many interstellar clouds are revealed by the recent {\em Herschel} observations \citep{Andre2013}, and the interaction of the cloud cores is a natural consequence of fragmentation of the filamentary cloud \citep{Inutsuka1997}.  

\subsection{Possible Site of Multiple Star Formation}

All of these facts above indicate that there are many regions with different evolutionary states even just within $\sim$1000 AU. 
There is a very low-luminosity source with a very compact outflow, and there are at least two high-density cores detected both in dust continuum emission and H$^{13}$CO$^{+}$($J$ = 3--2), indicating a possible formation site of multiple stars with separations of a few hundred AU.  There are also a few lower-density cores as shown in the HCO$^{+}$($J$ = 3--2) data.  Some gas components are dynamically interacting as seen in HCO$^{+}$($J$ = 3--2), and the compact outflow has an indication of interaction with the surrounding gas. 
These complex spatial and velocity structures indicate that the initial condition of low-mass star formation is highly dynamical.
The dynamical interaction of the gas may be also important for the determination of the stellar mass.
Actually, the lack of intensity enhancement in the observed lines toward the {\em Spitzer} source, which is shown in Section \ref{HDDC}, may be due to striping the surrounding gas or to the different velocity of the formed star compared with the surrounding gas in a turbulent environment.  The importance of this type of observational study is found to be the high spatial resolution observation with wide spatial frequency dynamic range.  Previous observations could not observe most of features detected here, and then the ALMA observations will revolutionary change the picture of star formation.

\acknowledgments

This Letter makes use of the following ALMA data: ADS/JAO.ALMA$\#$2011.0.00611.S. ALMA is a partnership of the ESO, NSF, NINS, NRC, NSC, and ASIAA. The Joint ALMA Observatory is operated by the ESO, AUI/NRAO, and NAOJ.  This work was financially supported by the Japan Society for the Promotion of Science (JSPS) KAKENHI grant Nos. 22244014, 23403001, and 23540270.  
K. Tomida is supported by JSPS Research Fellowship for Young Scientists.\\

\clearpage

\clearpage

\begin{deluxetable}{lcccccccccc}
\tabletypesize{\scriptsize}
\rotate
\tablecaption{Derived Parameters of Dense Cores \label{table1}}
\tablewidth{0pt}
\tablehead{
\colhead{Source} & \colhead{ $\alpha$} & \colhead{$\delta$} & \colhead{$F_{\rm \nu}$$^{\rm a}$} & \colhead{Size$^{\rm b}$} &
\colhead{$F_{\rm max}$$^{\rm c}$} & \colhead{$N_{\rm max}$$^{\rm c}$} & \colhead{Mass$^{\rm c}$} &
\colhead{$L_{\rm total}$$^{\rm d}$} & \colhead{$\Delta V$$^{\rm e}$} \\ \vspace{-0.7cm}\\
& ($J$2000) & ($J$2000) &\colhead{(mJy)} & \colhead{(AU)} & \colhead{(mJy beam$^{-1}$)} & \colhead{(cm$^{-2}$)} & \colhead{($M_\sun$)} & \colhead{(Jy beam$^{-1}$ km s$^{-1}$ arcsec$^2$)} & \colhead{(km s$^{-1}$)}}
\startdata
MMS-1 & 4$^{\rm h}$28$^{\rm m}$38\fs96 & +26\arcdeg51\arcsec35\farcs0 & 1.7 & 200 & 2.0 & \nodata & \nodata & \nodata & \nodata \\
             &                                                               &                                                             &         &        & \nodata   & \nodata & \nodata &\\
MMS-2 & 4$^{\rm h}$28$^{\rm m}$38\fs89 & +26\arcdeg51\arcsec33\farcs9 & 4.4 & 380 & 1.3 & 1.0 $\times$ 10$^{23}$ &3.5 $\times$ 10$^{-3}$ & 0.89 & 0.36\\
              &                                                               &                                                             &        &         & 3.7 $\times$ 10$^{2}$  & 3.7 $\times$ 10$^{22}$ & 1.5 $\times$ 10$^{-3}$\\
MMS-3 & 4$^{\rm h}$28$^{\rm m}$03\fs73 & +26\arcdeg51\arcsec38\farcs0 & 0.5 & 160 & 0.6 & 4.5 $\times$ 10$^{22}$ & 4.0 $\times$ 10$^{-4}$ & 0.27 & 0.40\\
              &                                                               &                                                             &        &         & 2.0 $\times$ 10$^{2}$  & 1.5 $\times$ 10$^{22}$ & 4.2 $\times$ 10$^{-4}$\\
\enddata
\tablecomments{$^{\rm a}$Flux of the dust emission integrated above the 3$\sigma$ level (1$\sigma$rms = 0.12 mJy beam$^{-1}$). \\
$^{\rm b}$Diameter of a circle having the same area above the 3$\sigma$ level. \\
$^{\rm c}$The first and second rows are evaluated from dust emission and H$^{13}$CO$^{+}$ ($J$ = 3--2) emission, respectively.\\
$^{\rm d}$Flux of H$^{13}$CO$^{+}$ ($J$ = 3--2) emission integrated above the 3$\sigma$ level of velocity-integrated intensity (1$\sigma$rms = 16.8 mJy beam$^{-1}$ km s$^{-1}$).\\
$^{\rm e}$Observed line width (FWHM) of spectrum of H$^{13}$CO$^{+}$($J$ = 3--2) averaged in each core by fitting the profile with a single Gaussian profile.\\
}
\end{deluxetable}

\end{document}